\documentclass{INTERSPEECH2023}

\interspeechcameraready

\title{Text-Only Domain Adaptation for End-to-End Speech Recognition through Down-Sampling Acoustic Representation}
\name{Jiaxu Zhu$^{1,3,\ddagger}$\thanks{$\ddagger$ Work done during internship at Tencent Inc.}, Weinan Tong$^{1,\dagger}$\thanks{$\dagger$ Equal contribution.  * Corresponding authors.}, Yaoxun Xu$^1$, Changhe Song$^{1,2}$, Zhiyong Wu$^{1,2,5,*}$, \\Zhao You$^{3,*}$, Dan Su$^3$, Dong Yu$^4$, Helen Meng$^5$}
\address{
  $^1$Shenzhen International Graduate School, Tsinghua University, Shenzhen, China\\
  $^2$Peng Cheng Lab, Shenzhen, China \\
  $^3$Tencent AI Lab, Shenzhen, China \quad
  $^4$Tencent AI Lab, Bellevue, WA, USA \\
  $^5$The Chinese University of Hong Kong, Hong Kong SAR, China}
\email{\{zhu-jx21, twn21, xuyx22, sch19\}@mails.tsinghua.edu.cn, zywu@sz.tsinghua.edu.cn, \{dennisyou, dansu, dyu\}@tencent.com, hmmeng@se.cuhk.edu.hk}

\begin{document}

\maketitle

\begin{abstract}
Mapping two modalities, speech and text, into a shared representation space, is a research topic of using text-only data to improve end-to-end automatic speech recognition (ASR) performance in new domains. However, the length of speech representation and text representation is inconsistent. Although the previous method up-samples the text representation to align with acoustic modality, it may not match the expected actual duration. In this paper, we proposed novel representations match strategy through down-sampling acoustic representation to align with text modality. By introducing a continuous integrate-and-fire (CIF) module generating acoustic representations consistent with token length, our ASR model can learn unified representations from both modalities better, allowing for domain adaptation using text-only data of the target domain. Experiment results of new domain data demonstrate the effectiveness of the proposed method.

\end{abstract}

\section{Introduction}

Automatic speech recognition (ASR) is a technology that converts audio into text. 
In recent years, end-to-end (E2E) ASR has attracted much attention and made great progress. E2E ASR can convert audio to text using a single network model, simplifying the training and inference process. There are three main types of E2E ASR models: connectionist temporal classification (CTC) \cite{graves2006connectionist}, recurrent neural network transducer (RNN-T) \cite{graves2012sequence, wang2019exploring}, and attention-based encoder-decoder (AED) \cite{chorowski2014end, chan2016listen}. 
Training with a large number of labeled data, the E2E ASR model has achieved excellent results.
However, the performance still has a serious decline in new domains. While the E2E ASR model requires paired audio-text for training, it is expensive to acquire high-quality paired labeled data for new domains.

Even more, due to the training paradigm with paired audio-text data, it is difficult for E2E ASR to directly use text-only data for domain adaptation like the traditional hidden Markov model (HMM)-deep neural network (DNN) hybrid speech recognition model.

Considering that the acquisition of unpaired text is relatively more convenient and the amount of text data is large, many studies attempted to leverage text-only data to adapt the E2E ASR model in new domains. 
A common approach is to use an external language model. This external language model uses a large amount of new domain unpaired text for training and fuses the E2E ASR model through the method of shallow fusion \cite{kannan2018analysis} or rescoring \cite{sainath21_interspeech}, to improve the recognition performance in the new domains.
Another approach is to generate audio from large amounts of text in the new domains through a text-to-speech (TTS) synthesis model \cite{chen2021injecting,chen2022tts4pretrain}, thus forming paired audio-text data that can be used for training E2E ASR models. However, this approach requires a reliable multi-speaker TTS model and the high computational cost of generating speech. Even worse, it still exists a mismatch between real and synthetic audio, which will affect the performance of speech recognition.

An alternative approach focuses on mapping the two modalities, speech and text, into shared representation spaces so that E2E ASR can be trained using paired audio-text or unpaired text \cite{ chen22r_interspeech, sainath2023joist}.
The main challenge of matching two modalities is that the length of speech representation and text representation is inconsistent, which makes it difficult for a model to learn a better-shared representation space. Although the previous approach aims to match the acoustic representation by replicating each text unit representation several times to increase the length of text representation or using a duration model to estimate phoneme and word alignments for each word in the transcript, it may not match the 
expected actual durations
in practice \cite{sainath2023joist,9747555}, which would affect the learning of shared representation space of two modalities.  
On the contrary, it will be more accurate to match the text length by down-sampling the acoustic representation \cite{dong2020cif}. 
We believe that modal matching with a more consistent length of representation will get better results.

In this paper, to explore the reasonable schemes of using text-only data for domain adaptation, we propose a new strategy to learn a shared representation space for the two modalities - speech and text, which can adapt E2E ASR to new domains more easily and effectively with text-only data.
To solve the problem of the inconsistent length of acoustic representation and text representation, 
inspired by \cite{dong2020cif} and \cite{gao22b_interspeech}, we introduce a continuous integrate-and-fire (CIF) module to generate the acoustic representations consistent with token length. Furthermore, considering that syllable is more pronunciation-related than character, which can be more effectively matched with acoustic representation, and syllable is more robust than character which can reduce the impact of rare word or long-tail word, we explore using syllable instead of character to generate a shared representation similar to acoustic representation. Together with a transformer-based syllable encoder, our ASR model can learn unified representations from both modalities better. Experiment results for out-of-domain data show that the proposed text-only domain adaptation performs well.

\begin{figure*}[ht]
\center{\includegraphics[width=0.85\linewidth]  {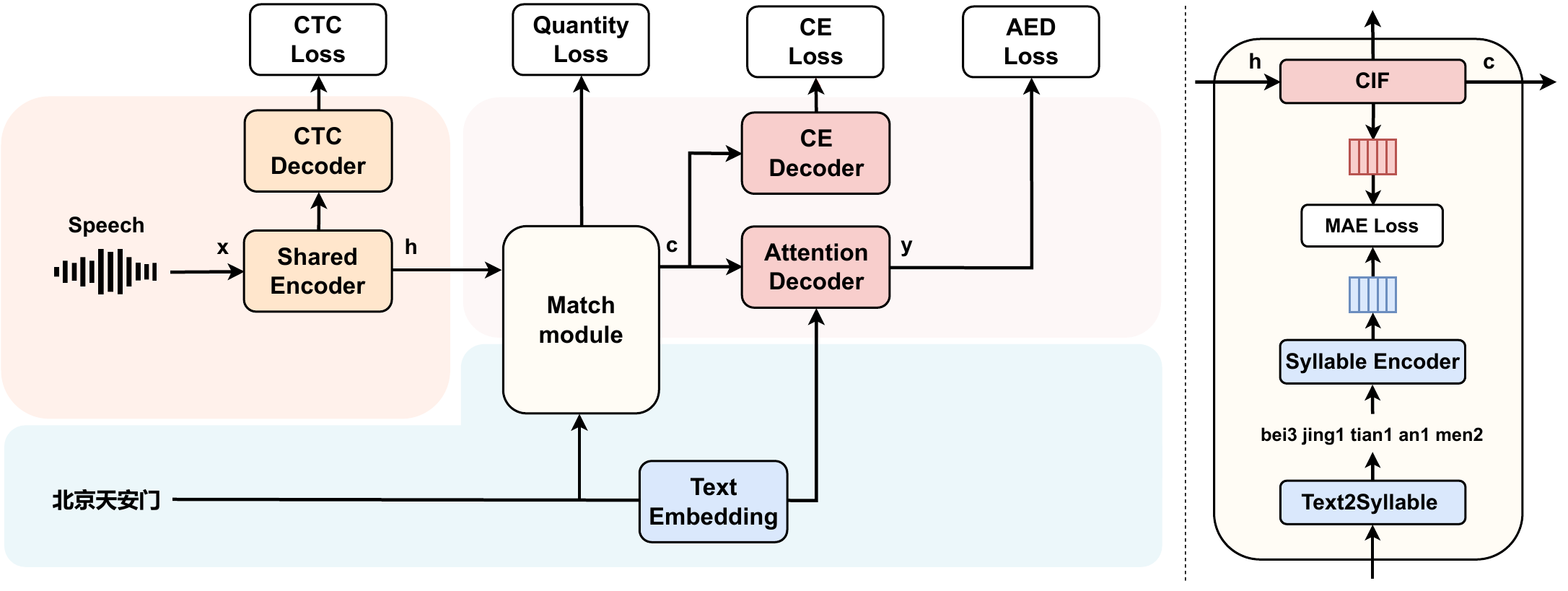}} 
\caption{\label{encoder} Proposed architecture of our E2E ASR model to learn unified representation from two modalities of speech and text. Left: the overall network; Right: the detailed architecture of the Match module}
\label{BaselineASR}
\end{figure*}

\section{Methodology}
In this section, we will review the architecture of our proposed method, which aims to apply a novel scheme to improve the ability to use text-only data for the E2E ASR model.
\subsection{Architecture}

Our proposed E2E ASR model is based on the one presented in WeNet \cite{yao21_interspeech}, which uses both CTC and Attention-based Encoder-Decoder (AED) losses during training to speed convergence and increase the AED model's robustness. 
As depicted in Figure \ref{BaselineASR}, the proposed ASR model mainly contains five parts, Attention-based \textit{Shared Encoder}, \textit{CTC Decoder}, \textit{CE Decoder}, \textit{Attention Decoder} and \textit{Match Module}.
The \textit{Shared Encoder} mainly consists of the Conformer \cite{gulati20_interspeech} blocks.
The \textit{CTC Decoder} and \textit{CE Decoder} consist of a linear layer and a log softmax layer. 
The \textit{Attention Decoder} mainly consists of multiple transformer \cite{VaswaniSPUJGKP17} blocks.
The \textit{Match Module} mainly consists of a CIF module and a transformer-based \textit{Syllable Encoder}.
Given that the module structure of \textit{CTC Decoder} and \textit{CE Decoder} is relatively simple, we mainly introduce other modules.

\subsubsection{The Shared Encoder}
The \textit{Shared Encoder} consists of a convolution subsampling layer containing two convolutional layers with stride 2 for downsampling, a linear projection layer, and a positional encoding layer, followed by multiple Conformer encoder layers. 
The \textit{Shared Encoder} transforms a $T$\text {-length } speech feature sequence $\mathbf{x}=\left(x_{1}, \ldots, x_{T}\right)$ to a $L$\text{-length} intermediate representation $\mathbf{h}=\left(h_{1}, \ldots, h_{L}\right)$, where $L \leq T $ owing to downsampling.

\subsubsection{The Match Module}
As depicted in Figure \ref{BaselineASR}, the \textit{Match Module} mainly consists of a CIF module and a transformer-based  \textit{Syllable Encoder}.
The CIF module consists of a 1-dimensional convolution layer and a linear layer to achieve a soft monotonic alignment.
The CIF encodes the \textit{Shared Encoder}'s outputs $\mathbf{h}=\left(h_{1}, \ldots, h_{L}\right)$ to predict the corresponding float weights $\mathbf{a}=(a_1,...,a_L)$ ranging from 0 to 1. 
We then carry out a weighted sum between $\mathbf{h}$ and $\mathbf{a}$ until the accumulated weight reaches a threshold which means reaching an acoustic boundary and generating a new integrated embedding. The threshold is recommended to be 1.0 in \cite{dong2020cif}. In this way, CIF outputs high-level acoustic sequence $\mathbf{c} = (c_1,...,c_I )$, which is consistent with the length of the text representation. 
On the other hand, a transformer-based  \textit{Syllable Encoder} takes syllable embedding as inputs and output text representations.
Considering that the acoustic sequence $\mathbf{c}$ is strictly aligned with the text sequence during training, we map the two modalities into a shared space with a mean absolute error (MAE) training objective.

\subsubsection{The Attention Decoder}
The \textit{Attention Decoder} consists of a positional encoding layer, multiple transformer decoder layers, and a linear projection layer.
Given $\mathbf{c}$ and previously emitted character outputs 
$\mathbf{y}_{0: i-1}=\left(y_{0}, \ldots, y_{i-1}\right)$, the attention decoder predictes the next character $y_i$.

\subsection{The Overall Training Pipeline}
In this subsection, we describe the overall modality-matched training process using available paired speech-text data and the text-only training process using unpaired text-only data. Our method aims to utilize a large number of unpaired text data without modifying the model. To do so, we allow the model to be trained on either paired audio-text data or unpaired text data.

\subsubsection{Modality Matched Training}
To solve the actually expected duration mismatch of text units in previous approaches, we have changed a perspective to down-sample the acoustic representation instead of up-sampling text representation. As shown in Figure \ref{BaselineASR}, speech is first transformed into a long acoustic representation by the \textit{Shared Encoder}. Then, in the \textit{Match Module}, a CIF module down-samples the long acoustic representation to a  relatively short acoustic representation with the same length as the text representation generated by the \textit{Syllable Encoder}.
Although we use syllables as model unit to get a hidden representation, we still use the word ``text representation'' to refer to it, because the syllables come from text through a text2syllable process which mainly uses an open-source Chinese character to pinyin tool python-pinyin$\footnote{https://pypi.org/project/pypinyin/}$. 

\subsubsection{Text-only Training}
The \textit{Attention Decoder} in our E2E ASR model is a transformer-based decoder that mainly consists of a self-attention module, a cross-attention module, and a feed-forward module. The self-attention module allows the decoder to model the content text between token sequences \cite{deng2021improving}. Therefore, the \textit{Attention Decoder} can interpret as an internal language in E2E ASR \cite{9383515,9415039,zeineldeen21_interspeech}.
However, the cross-attention module makes the decoder dependent on the acoustic encoder output and thus can not be separately trained on text-only data, which also makes the internal language model hard to update for domain adaptation. And the approach of shared representation space of speech and text would be an effective way to use text-only data, in which the text representation plays a similar role to acoustic representation.

\begin{figure}[ht]
\center{\includegraphics[width=0.85\linewidth]  {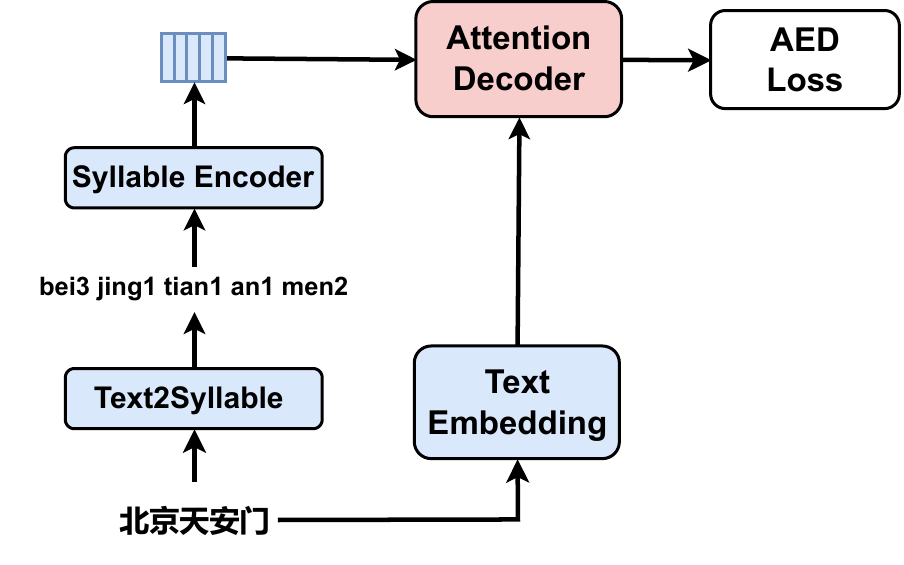}} 
\caption{\label{encoder} Overview of text-only training process.}
\label{text-only}
\end{figure}

After the paired speech-text data training process, the \textit{Syllable Encoder} can transform syllable embedding to a hidden representation, which shares a representation space with acoustic representation. Therefore, we can utilize several text-only data to train the E2E ASR model. The detailed process is as depicted in Figure \ref{text-only}, we first get a hidden representation from \textit{Syllable Encoder} instead of CIF to replace the absent acoustic representation in the text-only process. And then we only train the \textit{Attention Decoder} while other module parameters are fixed to not update. The \textit{Attention Decoder} is based on transformer architecture and it does not require changing the original objective function. 
Considering that the internal language model has some parts related to acoustic modeling, it is not a real language model. In order to prevent the decoder from affecting the acoustic modeling part and suppress catastrophic forgetting during text-only training, we randomly use audio-text data of the source domain in training to ensure that the decoder can perform ASR tasks when conditioned on audio features \cite{wang21t_interspeech,kim-etal-2022-joint}.
Therefore, in our novel modality-matched schemes, we can implement the text-only training process more simply and effectively.

\subsection{Loss Function}
We used five loss functions to train our model, namely the CTC, Quantity, CE, AED, and MAE losses, where the CTC, Quantity, and CE loss just like that in \cite{dong2020cif}. The types are jointly trained, as follows:
\begin{equation}
\mathcal{L}=\alpha\mathcal{L}_{CTC} + \beta\mathcal{L}_{QUA} + \gamma\mathcal{L}_{CE} + \lambda\mathcal{L}_{AED} + \delta\mathcal{L}_{MAE}
\end{equation}
Where $\alpha, \beta, \gamma, \lambda$ and $\delta$ are tunable parameters. In the experiment, we set $\alpha$ and $\gamma$ to 0.5, and other parameters to 1.

In text-only training, AED loss is the only one used:
\begin{equation}
\mathcal{L}_{text}= \mathcal{L}_{AED} 
\end{equation}

\section{Experiments}
\subsection{Datasets}
In this paper, we train our proposed E2E ASR on public Mandarin Aishell-1 \cite{bu2017aishell} datasets.
The Aishell-1 corpus consists of 178 hours of labeled speech collected from 400 speakers. The content of the datasets covers 5 domains including Finance, Science and Technology, Sports, Entertainment, and News. To compare the domain adaptation ability
of ASR in the text domain while minimizing the influence of differences in the acoustic environment, we chose another public Mandarin dataset Aishell-2 \cite{du2018aishell} that has a similar acoustic environment for sound recording but the corresponding text contents cover different text domains.
The Aishell-2 corpus consists of 1000 hours of labeled speech collected from 1991 speakers. 
The content of Aishell-2 correspondent-only domains of voice commands, digital sequence, places of interest, entertainment, finance, technology, sports, English spellings, and free speaking without specific topics.
Furthermore, we also conducted further experiments on different
domains on the WenetSpeech \cite{zhang2022wenetspeech}, which is a multi-domain Mandarin corpus consisting of high-quality labeled speech but a relatively more complex acoustic environment than Aishell-1.
We use the Aishell-1 training set for training and the development set for early stopping. 

\subsection{Experimental Setup}
For all experiments, we use the open-source WeNet toolkit \cite{yao21_interspeech} to build our proposed ASR model. 
And we used the default values in the WeNet for the main parameters which have been validated by the WeNet contributor.
The input features are 80-dimensional log Mel-filterbank (FBank) computed on a 25ms window with a 10ms shift.
We use SpecAugment \cite{park19e_interspeech} and speed perturb for data augmentation.
We choose 4233 characters (including 〈blank〉, 〈unk〉, 〈sos/eos〉 labels) as model units for Aishell-1.

Following the WeNet recipe \cite{yao21_interspeech}, we construct the base model using 12 Conformer blocks in the \textit{Shared Encoder}, 6 transformer blocks in the \textit{Attention Decoder} and 4 transformer blocks in the \textit{Syllable Encoder}.
We employ $h$ = 4 parallel attention heads in both the Conformer block and transformer block.
For every layer, we use $d_k = d_v = d_{model}/h$ = 64, $d_{ffn}$ = 2048. 

We train the model with Adam Optimizer \cite{VaswaniSPUJGKP17} for at most 240 epochs with 12 batches.
And \textit{learning rate} = 0.002, \textit{warm up} = 25000, and gradient clipping at 5.0. 
Additionally, during training, we employ the gradient accumulation method, in which the gradients are modified every four batches.
Moreover, we employ label smoothing of value $\epsilon_{l s}$ = 0.1 and a dropout rate of $P_{drop}$ = 0.1. 
We set the weight $\lambda$ of the CTC branch during joint training to 0.3. 
We also train the n-gram language model with new domains of text-only data follow by the WeNet recipe.
During joint decoding, we set the CTC-weight $\lambda$ to 0.5.
To avoid overfitting, we averaged the 30 best model parameters in the development dataset.

\section{Results}

The performance of the models is evaluated based on character error rates (CER). 
Our experimental results are mainly based on the attention-rescore two-step decoding method.

\subsection{Main Result}
Our method is evaluated on the Aishell-1 dataset. We compare the proposed ASR model with other models in the literature. As shown in Table \ref{main_res}, the proposed ASR model achieves comparable performance with a series of state-of-the-art approaches.
\begin{table}[h]
  \centering
  \caption{Main results on Aishell-1 (CER)}
    \begin{tabular}{ccc}
    \toprule
    \textbf{Model} & \textbf{dev} & \textbf{test} \\
    \hline  Espnet Conformer \cite{watanabe18_interspeech} & 4.5 & 4.9 \\
    \hline WeNet Conformer \cite{yao21_interspeech} & - & 4.6  \\
    \hline Branchformer \cite{peng2022branchformer} & 4.2 & 4.4  \\
    \hline Blockformer \cite{ren2022improving} & - & 4.4  \\
    \hline CIF-based model \cite{dong2020cif} & 4.5 & 4.9  \\
     + CE Decoder & 4.2 & 4.7 \\
     +CE Decoder and Match module (Proposed) &  4.1 & 4.5 \\
    \hline
    \end{tabular}
  \label{main_res}
\end{table}



\subsection{Domain Adaption}
In order to prove the effectiveness of our text-only method in domain adaptation, we also compare the results on the Aishell-2 test and dev datasets, which have a similar acoustic environment with Aishell-1 but cover different text domains. We use the text data of the Aishell-2 training dataset for text-only training. As shown in Table 2, after text-only training, the performance of the ASR model in a new domain is significantly improved. The LM is trained with Aishell-2 text data  from the training set. And  text-only refer to the model after text-only training.



\begin{table}[h]
  \centering
  \caption{Comparison of the performance after text-only domain adaption on AishellL-2 (CER)}
    \begin{tabular}{ccc}
    \toprule
    \textbf{Model} & \textbf{Aishell-2 dev} & \textbf{Aishell-2 test} \\
    \hline  Proposed Model  & 11.7 & 11.6 \\
    \hline  + LM  & 11.5 & 11.4 \\
    \hline Text-Only & \textbf{11.2} & \textbf{11.0}  \\
    \hline
    \end{tabular}
  \label{aishell-2}
\end{table}


In addition, to further illustrate the validity of our approach in more difficult text domains and more complex acoustic environments, we conduct further experiments on three domains of the WenetSpeech dataset. As shown in Table \ref{wenetspeech}, the proposed text-only method can also improve recognition performance. However, with the increasing complexity of data, the recognition performance of the proposed model is poor, and the performance of text-only is not obvious. Our proposed model is trained on a relatively small and quiet dataset, and its ability for acoustic modeling is not strong enough, so the performance is poor in a dataset with a complex acoustic environment.

\begin{table}[h]
  \centering
  \caption{Comparison of the performance after text-only domain adaption on WeNetSpeech (CER)}
    \begin{tabular}{cccc}
  \toprule
    \textbf{Model} & \textbf{audiobook} & \textbf{interview} & \textbf{drama}\\
    \hline  Proposed Model  & 15.0  & 37.9 & 56.6 \\
    \hline Text-Only  & \textbf{14.2}  & 37.9 & \textbf{56.4} \\
    \hline
    \end{tabular}
  \label{wenetspeech}
  \vspace{-0.3cm}
\end{table}




\subsection{Effects of the Epoch of Text-Only}
\begin{figure}[h]
\center{\includegraphics[width=0.95\linewidth]  {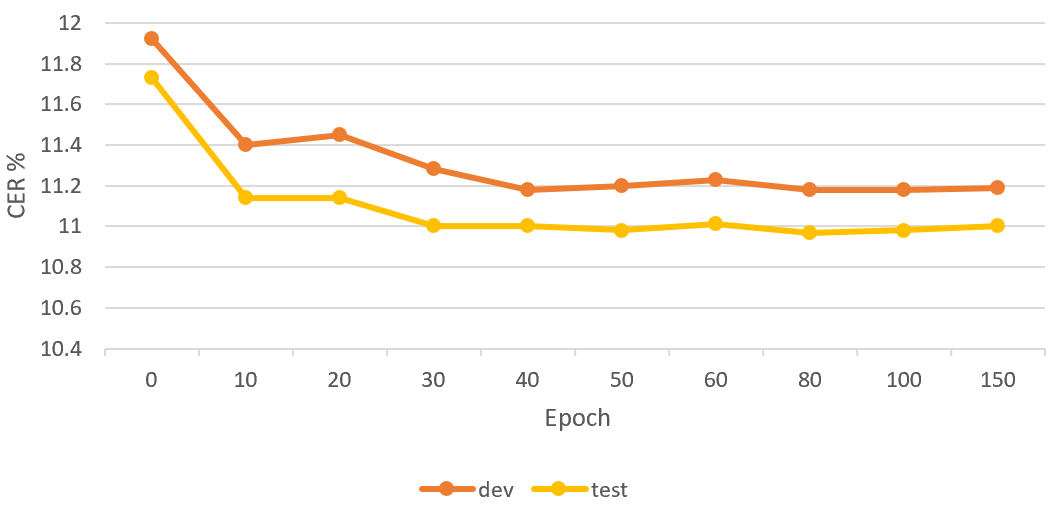}} 
\caption{\label{epoch} Text-only performance changes with epoch.} 
\end{figure}

We further study the out-of-domain performance changed with the training epoch. As shown in Figure \ref{epoch}, in the beginning, with the increase of training epochs, the ASR performance was significantly improved. When the training epoch reaches nearly 40, the text-only performance will be achieved. This shows that the text-only method can achieve domain adaptation quickly.

\subsection{Analysis on Different Model Unit for Text-Only}
Furthermore, we also study the performance of the different model units of character and syllable in our text-only. As shown in Table \ref{units}, using syllables as model unit achieves a better performance than using the character in the proposed model and is more effective in text-only domain adaptation. On the one hand, the syllable modeling units are more pronunciation-related than character, which can be more effectively matched with acoustic representation. On the other hand, the rare character or long-tail character may be difficult to fully model in a modality matching. 

\begin{table}[h]
  \centering
  \caption{Comparison of the performance of different model unit in text-only domain adaption on AishellL-2 (CER)}
    \begin{tabular}{ccc}
    \toprule
    & \textbf{Character} & \textbf{Syllable} \\
    \textbf{Model} & \textbf{dev} / \textbf{test} & \textbf{dev} / \textbf{test} \\
    \hline  
    Proposed Model  & 11.8 / 11.7 & 11.7 / 11.6 \\
    + LM & \textbf{11.6} / \textbf{11.5} & 11.5 / 11.4  \\
    Text-Only & 12.0 / 11.9 & \textbf{11.2} / \textbf{11.0}  \\
    \hline
    \end{tabular}
  \label{units}
  \vspace{-0.2cm}
\end{table}

\section{Conclusions}
In this paper, we proposed a novel representations match module through down-sampling acoustic representation to align with text modality. 
By introducing a continuous integrate-and-fire (CIF) module generating acoustic representations consistent with token length and using pronunciation-related model unit syllable matching acoustic representation effectively, our ASR model can learn unified representations from both modalities better, allowing for domain adaptation using text-only data of the target domain. Experimental comparisons for out-of-domain settings demonstrate that the proposed text-only domain adaptation achieves a good performance. In the future, we will further explore the performance of different model units with large-scale datasets and verify the performance of our method on the English datasets.

\section{Acknowledgements}
This work is supported by Shenzhen Science and Technology Program (WDZC20200818121348001, JCYJ20220818101014 030), the Major Key Project of PCL (PCL2021A06, PCL2022 D01) and AMiner.Shenzhen SciBrain fund.


\bibliographystyle{IEEEtran}
\bibliography{mybib}

\end{document}